\documentclass[twocolumn,tighten]{aastex63}

\usepackage{times,natbib,graphicx,amsmath,multirow,xspace}
\usepackage{xcolor}
\usepackage{lineno}
\usepackage{hyperref}
\newcommand{\xrism}{XRISM\xspace}
\newcommand{\nustar}{NuSTAR\xspace}

\newcommand{\chandra}{Chandra\xspace}

\newcommand{\ms}{$M_{\odot}$\xspace}

\newcommand{\rin}{$R_{\rm in}$\xspace}
\newcommand{\rg}{$R_{g}$\xspace}
\newcommand{\risco}{$R_{\mathrm{ISCO}}$\xspace}

\newcommand{\relxillns}{{\sc relxillNS}\xspace}

\newcommand{\source}{\mbox{Serpens X-1}\xspace}

\shorttitle{Resolving Serpens X-1}
\shortauthors{Ludlam et al.}

\begin{document}

\title{\xrism Resolves Relativistic Effects from the Innermost Accretion Disk in Serpens X-1}

\correspondingauthor{R.~M.~Ludlam}
\email{ef2051@wayne.edu}

\author[0000-0002-8961-939X]{R.~M.~Ludlam}
\affiliation{Department of Physics \& Astronomy, Wayne State University, 666 West Hancock Street, Detroit, MI 48201, USA}
\author[0000-0003-2869-7682]{J. M. Miller}
\affiliation{Department of Astronomy, University of Michigan, 1085 South University Ave, Ann Arbor, MI 48109-1107, USA}
\author[0000-0002-8294-9281]{E.~M.~Cackett}
\affiliation{Department of Physics \& Astronomy, Wayne State University, 666 West Hancock Street, Detroit, MI 48201, USA}
\author[0000-0003-3828-2448]{J.~A.~Garc\'{i}a}
\affiliation{Cahill Center for Astronomy and Astrophysics, California Institute of Technology, 1200 E. California Blvd, MC 290-17, Pasadena, CA, 91125, USA}

\begin{abstract}

We present the first \xrism/Resolve observation of the persistently accreting neutron star (NS) low-mass X-ray binary Serpens X-1. The source was observed on October 17th, 2024, for approximately 350 ks of elapsed time, resulting in 171 ks of exposure. The source exhibited 22\% variability with respect to the average count rate of 73.1 count/s during the observation, but remained in a spectrally soft state throughout. The time averaged spectrum was analyzed in conjunction with spectra extracted from periods of different count rate to check for variations in spectral components. The unprecedented energy resolution of 4.5 eV at 6 keV of \xrism/Resolve provides a detailed look at the shape and structure of the Fe emission line within the data, which shows a dual-peaked structure with an extended red-wing, and steep decline in the blue-wing of the line profile.  Fits with the reflection model \relxillns are able to describe the structure in the Fe line region, and confirms previous results that the disk is close to the NS (\rin = $1.02_{-0.01}^{+0.21}$ \risco).  These models also measure a low systemic inclination ($i=5^{\circ}\pm1^{\circ}$), confirming prior X-ray and optical studies. Alternative models were explored to describe the structure of the Fe line profile, however, relativistic reflection provides the simplest and statistically best explanation of the data.    
\end{abstract}


\section{Introduction} \label{sec:intro}
Low-mass X-ray binaries (LMXBs) consist of a stellar mass companion of $\lesssim1\ M_{\odot}$ that transfers material onto a compact object via Roche-lobe overflow. 
Bright, persistent neutron stars (NSs) in LMXBs are divided into two categories based upon X-ray spectral properties and variability: \lq \lq atoll" and \lq \lq Z" type \citep{hasinger89}. The two classes acquire their name from the tracks that the sources trace out in the hardness-intensity diagram (HID) and color-color diagram (CCD). Atolls are less luminous in comparison to Z sources \citep{vanderklis05} and have two distinct spectral states: the hard ``island" state and soft ``banana" state. The harder spectral state is dominated by Comptonized emission from the corona region with an additional thermal component with a temperature $\lesssim1$ keV \citep{barret00, church01}, whereas the softer spectral state is dominated by thermal emission with weak Comptonization. There are a number of models used to account for the observed emission in the soft state, such as a single-temperature blackbody and Comptonized disk (``Western" model: \citealt{white88}), a disk blackbody with a Comptonized single-temperature blackbody (``Eastern" model: \citealt{mitsuda89}), and more recently the ``hybrid" model \citep{Lin07} that contains a disk blackbody, single-temperature blackbody, and power-law component.
Variations of these models are used in the literature when fitting soft state spectra of NS LMXBs (see \citealt{ludlam23} for a review and further references therein).

\begin{figure*}[t!]
 \begin{center}
  \includegraphics[width=0.9\textwidth]{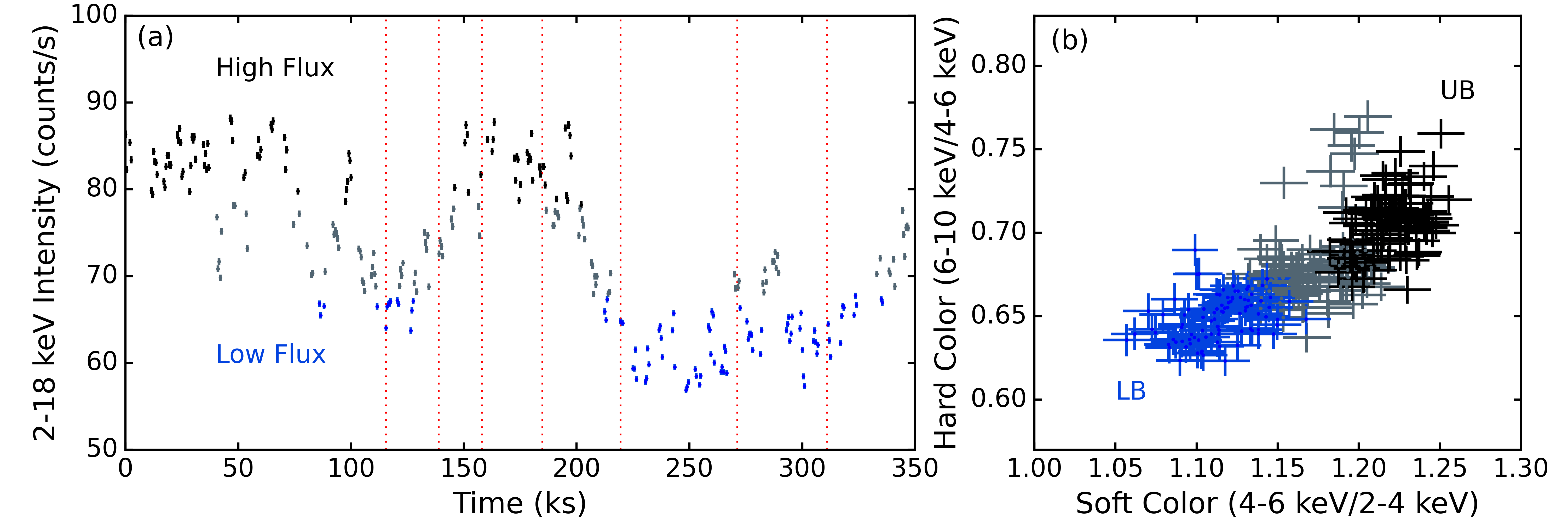} 
 \end{center}
\caption{The \xrism/Resolve (a) light curve and (b) color-color diagram for \source. The average count rate during the observation was 73.1 counts/s with variability of 22\% from the average. We designate the high flux state as count rates in excess of 78.2 counts/s (upper 15\% of the source intensity) and the low flux state as count rates below 67.9 counts/s. The vertical dotted lines indicate the time of the bursts that were removed from the data analyzed herein. The color-color diagram shows that the source remains in the soft banana state throughout the observation. The designated low and high flux states correspond to the lower banana (LB) and upper banana (UB) states, respectively. The \xrism data were binned to 500 s. 
}\label{fig:lc}
\end{figure*}

When the accretion disk surrounding the NS is externally illuminated by X-rays originating from close to the compact object, such as the corona \citep{Sunyaev1991} or boundary/spreading layer \citep{popham2001, inogamov99}, the photons are reprocessed and re-emitted as a series of atomic features superimposed onto a `reflected' continuum. This component is collectively referred to as the reflection spectrum \citep{Ross2005, Garcia2010}. When these features arise close to the NS, they are shaped by relativistic Doppler shifts and gravitational red-shifts \citep{Fabian1989, Dauser2013}. The strength of each effect depends on the proximity to the NS (e.g., stronger gravitational redshift closer to the compact object: \citealt{Fabian00}) and the systemic orientation with respect to our line of sight (e.g., stronger Doppler broadening at higher inclinations: \citealt{Dauser10}). The most prominent emission line in the reflection spectrum is the iron (Fe) K-shell transition, which exhibits an asymmetric line profile shaped by the effects of the inner disk. 
Therefore, reflection features are often used to infer fundamental properties of the compact object (e.g., upper limit on NS radius or magnetic field strength: \citealt{Bhattacharyya07, cackett08, miller13, Ludlam17a}), as well as the accretion disk itself (e.g., location of the inner disk, inclination, density, etc: see \citealt{ludlam24} for further review and references therein).

\source is significant as it was the first NS LMXB discovered to exhibit a broad asymmetric Fe line indicative of  relativistically broadening \citep{Bhattacharyya07, cackett08}, which opened up the field of X-ray reflection spectroscopy in these systems.  It is a persistently accreting ``atoll'' source located at a distance of $7.7\pm0.9$~kpc away \citep{galloway2008}.  The source varies in flux at the 50\% level \citep{mondal20}, but is only observed in the soft `banana' state of the CCD and HID.  Within the soft (banana) state for atolls, the hardness ratio remains relatively constant \citep{church14}.  For these reasons, spectra that sample long periods are physically meaningful, and 
\source has been observed by every major X-ray mission \citep{Oosterbroek01,church01,Bhattacharyya07,cackett08,cackett10,Chiang16, miller13,matranga17,ludlam18, ursini2024b, hall25}.  

Optical spectroscopy has inferred an orbital period of just $P\simeq 2$~hours in \source, and an inclination of $i \leq 10^{\circ}$ \citep{Cornelisse13}.  This inclination is supported by some relativistic reflection modeling in X-rays, \citep{cackett10, miller13,ludlam18,hall25}, but other modeling suggests higher inclination values, $16^{\circ}<i<50^{\circ}$ \citep{matranga17,mondal20,cackett08, Chiang16, ursini2024b}. Additionally, there are conflicting measurements about the location of the inner accretion disk. For the same \nustar dataset (energy resolution of 400 eV at 6 keV), \cite{miller13} found a disk location consistent with the innermost stable circular orbit (ISCO; where 1\risco= 6 $GM/c^2$ = 6\rg for a dimensionless spin parameter of 0), while \cite{matranga17} reported a truncated disk at $>$2\risco. 

High resolution X-ray spectra provides a means to definitively reveal relativistic and Doppler effects on the shape of the red- and blue-wing of the Fe line profile.
The only previous spectrum of \source with better than 100 eV energy resolution was taken with \chandra/HETG \citep{Chiang16}. Though the source was observed for 300 ks of exposure, the signal was insufficient to discern sharp features within the Fe line profile.  The energy resolution of \xrism/Resolve (4.5 eV at 6 keV: \citealt{xrism}) is nearly $10\times$ better than \chandra/HETG, making it the ideal instrument for evaluating the origin of spectral lines. We present the first \xrism/Resolve observation of \source in order to investigate the structure of the Fe line region. The presentation of the paper is as follows: \S~2 discusses the reduction and handling of the data, \S~3 presents the analysis and results, and \S~4 discusses the results in the context of the literature.

\section{Observation and Data Reduction} \label{sec:data}
\xrism observed \source on October 17th, 2024 for approximately 171 ks of cumulative exposure (ObsID: 201073010). 
The soft X-ray imaging telescope Xtend \citep{xtend} was operated in the 1/8 window$+$burst mode. This instrument has an energy resolution of 170--180 eV at 6 keV. Given that the focus of this letter on the fine structure of the Fe line profile, we solely focus on the high energy resolution spectra provided by Resolve and defer analysis of Xtend data to a subsequent investigation.
Due to the gate valve being closed, no filter was necessary to reduce the illuminating flux of the source on the Resolve instrument. We follow the \xrism ABC Guide\footnote{https://heasarc.gsfc.nasa.gov/docs/xrism/analysis/abc\_guide/xrism\_abc.pdf} with publicly available CALDB version index 20250315 applied via {\sc xapipeline} to reduce the Resolve data with pixel 27 excluded. Inspection of the gain report for this observation did not reveal issues for any other pixel. Only the highest resolution events (``Hp'') were considered and extracted for light curves and spectra. There were seven Type-I X-ray bursts that occurred during the observation that were identified via inspection of the light curve binned to 10 second intervals. GTIs were created in mission elapsed time (MET) to remove 60--90 seconds intervals containing the burst emission (duration beginning 10 seconds prior to the burst until the emission returns to the persistent emission level). These were applied to the event file to remove bursts from the subsequent analysis as we are concerned with modeling the reprocessed features during the persistent emission. Light curves were extracted in the 2--4 keV, 4--6 keV, 6--10 keV, and 2--18 keV energy bands using 500 second time bins to create the color-color diagram and light curve presented in Figure \ref{fig:lc}. The source remains in the soft banana state throughout the observation.  The average array count rate was 73.1 count/s with 22\% variability with respect to the mean in the overall intensity during the observation. 

In addition to a time-averaged spectrum of the persistent flux, spectra were extracted from low-flux and high-flux states (i.e., intensity corresponding to the lower and upper 15\% of the source variability, respectively) to check for variations in inferred parameters between the lower banana (LB) and upper banana (UB) states.  The total exposure per spectrum after filtering is 171 ks, 49.6 ks, and 47.6 ks for the time-averaged, LB, and UB, respectively. Large response files were generated via {\sc rslmkrmf} and {\sc xaarfgen} for each spectrum. 
Background models for \xrism/Resolve are under development, but there is a provisional Resolve non-X-ray background (NXB\footnote{\href{https://heasarc.gsfc.nasa.gov/docs/xrism/analysis/nxb/resolve_nxb_db.html}{Resolve NXB database}}) model. We use this to estimate the signal-to-noise ratio (SNR) of the spectra in the energy band considered in our analysis presented in \S~3. Given that the count rate of the NXB is $\lesssim10^{-3}$ count/s, we find a SNR $> 100$ below 11 keV and $>10$ out to 17.4 keV for all extracted spectra; indicating a positive source detection out to high energies. 
The data were inspected for bright source effects such as event loss and electronic cross-talk \citep{ishisaki18, mizumoto25, 2025arXiv250606692M}, but are not an issue for the persistent emission given that the array count rate does not exceed 200 counts/s. 

\begin{figure}[t]
 \begin{center}
  \includegraphics[width=0.45\textwidth]{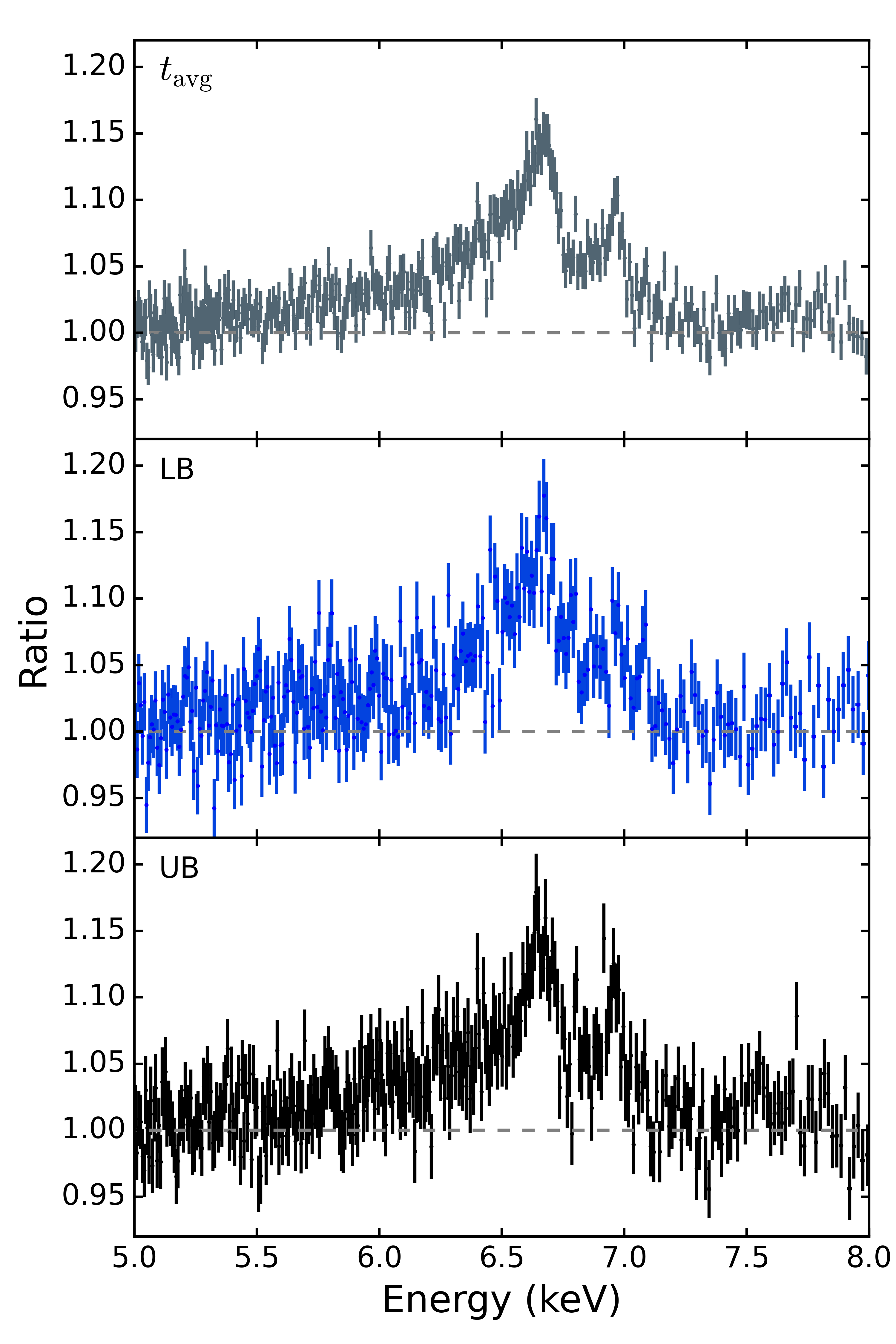} 
 \end{center}
\caption{The ratio of the \xrism/Resolve data to the continuum model for the time-averaged ($t_{\rm avg}$), lower banana (LB) state, and upper banana (UB) state, respectively. Data were rebinned for clarity. 
}\label{fig:fe}
\end{figure}

\section{Analysis and Results} \label{sec:results}

\begin{table*}[!t]

\caption{Spectral modeling \xrism/Resolve time-averaged ($t_{\rm avg}$), lower banana (LB), and upper banana (UB) data.}
\label{tab:spec} 

\begin{center}

\begin{tabular}{ll|ccc|ccc}
\hline

Model & Parameter & \multicolumn{3}{c|}{Continuum} & \multicolumn{3}{c}{Reflection}\\
& & $t_{\rm avg}$ & LB & UB & $t_{\rm avg}$ & LB & UB\\
\hline

{\sc diskbb} %%%%
& $kT_{\rm in}$ (keV) 
& $ 0.68 \pm0.01$
& $ 0.69 \pm0.02$
& $ 0.68 \pm0.02$
& $ 1.04 _{- 0.04 }^{+ 0.02 }$
& $ 1.02_{- 0.06 }^{+ 0.04 }$
& $ 1.07\pm0.05$
\\

& norm$_{\rm disk}$ 
& $ 785 _{- 46 }^{+ 49 }$
& $ 699 _{- 80 }^{+ 86 }$
& $ 800 _{- 88 }^{+ 100 }$
& $ 123_{- 15 }^{+ 12 }$
& $ 123\pm22$
& $ 121_{- 19 }^{+ 16}$
\\

{\sc bbody} %%%%
& $kT_{\rm bb}$ (keV) 
& $ 1.43 \pm0.01$
& $ 1.40 \pm0.01$
& $ 1.47 \pm0.01$
& ...
& ...
& ...
\\
& norm$_{\rm bb}$ ($10^{-2}$)
& $ 3.88 \pm0.08$
& $ 3.2 \pm0.1$
& $ 4.5 \pm0.2$
& ...
& ...
& ...
\\

{\sc powerlaw} 
& $\Gamma$
& $1.7\pm0.1$
& $1.8\pm0.2$
& $1.6\pm0.2$
& $ 2.13_{- 0.02 }^{+ 0.08 }$
& $ 2.2\pm0.1$
& $ 2.0\pm0.1$
\\
& norm$_{pl}$
&  $ 0.19 _{- 0.05 }^{+ 0.06 }$
&  $ 0.3 _{- 0.1 }^{+ 0.2 }$
&  $ 0.2 \pm0.1$
& $ 0.53 \pm0.08$
& $ 0.6 _{- 0.1 }^{+ 0.2 }$
& $ 0.4 _{- 0.1 }^{+ 0.2 }$
\\

{\sc relxillNS} %%%%
& $q$
& ...
& ...
& ...
&$ 2.04 _{- 0.07 }^{+ 0.05 }$
&$ 2.0 \pm0.1$
&$ 2.1 \pm0.1$
\\

& $i\ ^{\dagger}$ ($^{\circ}$) 
& ...
& ...
& ...
&$ 5 \pm1 $
&$ 5 \pm1 $
&$ 5 \pm1 $
\\

& $R_{\rm in}$ ($R_{\rm ISCO}$)
& ...
& ...
& ...
& $ 1.02 _{- 0.01 }^{+ 0.21 }$
&$ 1.04 _{- 0.03 }^{+ 0.40 }$
&$ 1.11 _{- 0.10 }^{+ 0.33 }$
\\

& $R_{\rm in}$ (\rg)
& ...
& ...
& ...
& $ 6.12 _{- 0.06 }^{+ 1.26 }$
&$ 6.24 _{- 0.18 }^{+2.40 }$
&$ 6.66 _{- 0.60 }^{+1.98 }$
\\

& $kT_{bb}$ (keV)
& ...
& ...
& ...
& $ 1.84 _{- 0.08 }^{+ 0.01 }$
&$ 1.80 _{- 0.11 }^{+ 0.04 }$
& $ 1.89 _{- 0.09 }^{+ 0.07 }$
\\

& $\log (\xi)$ 
& ...
& ...
& ...
& $ 3.40 \pm0.03$
&$ 3.39 \pm0.06$
&$ 3.43 _{- 0.04 }^{+ 0.07 }$
\\

& $A_{\rm Fe}\ ^{\dagger}$ 
& ...
& ...
& ... 
& $ 3.3 _{- 0.3 }^{+ 1.1 }$
& $ 3.3 _{- 0.3 }^{+ 1.1 }$
& $ 3.3 _{- 0.3 }^{+ 1.1 }$
\\

& $\log(n_{e}/\rm cm^{-3})\ ^{\dagger}$
& ...
& ...
& ... 
& $ 18.0 _{- 0.2 }^{+ 0.1 }$
& $ 18.0 _{- 0.2 }^{+ 0.1 }$
& $ 18.0 _{- 0.2 }^{+ 0.1 }$
\\

& $f_{\rm refl}$ 
& ...
& ...
& ...
&$ 0.61 _{- 0.16 }^{+ 0.04 }$
&$ 0.61 _{- 0.17 }^{+ 0.04 }$
&$ 0.56 _{- 0.14 }^{+ 0.04 }$
\\

& norm$_{\rm refl}$ ($10^{-3}$) 
& ...
& ...
& ...
&$ 1.88 _{- 0.08 }^{+ 0.23 }$
&$ 1.52  _{- 0.07 }^{+ 0.21 }$
&$ 2.3 _{- 0.1}^{+ 0.3 }$
\\

\multicolumn{2}{c|}{C-statistic (dof)} & 7548.60 (4773) & 5056.10 (4371) & 5120.66 (4430) & 5205.64 (4766) & 4522.03 (4367) & 4390.89 (4426)\\ 
\multicolumn{2}{c|}{BIC} & 7599.43 & 5106.40 & 5171.05 & 5315.78 & 4605.87 & 4474.87\\ 

\hline
\multicolumn{2}{c}{total C-statistic (dof)} & \multicolumn{3}{c}{17725.36 (13574)} & \multicolumn{3}{c}{ 14118.56 (13559)}\\ 
\hline

\multicolumn{6}{l}{$^{\dagger}=$ tied between all branches} 

\end{tabular}
\end{center}

\medskip
Note.---  Errors are reported at the 90\% confidence level. 
Interstellar absorption along the line of sight is modeled with {\sc tbabs} and fixed to $N_{\mathrm{H}}=4.4\times10^{21}$ cm$^{-2}$ \citep{HI4PI}.
The normalization of {\sc bbody} is defined as $(L/10^{39}\ \mathrm{erg\ s^{-1}})/(D/10\ \mathrm{kpc})^{2}$. 
The normalization of {\sc diskbb} is defined as $(R_{in}/\mathrm{km})^{2}/(D/10\ \mathrm{kpc})^{2}\times\cos{\theta}$. The power-law  normalization is defined as photons~keV$^{-1}$~cm$^{-2}$~s$^{-1}$ at 1~keV. For \relxillns, the dimensionless spin is fixed at $a=0$, such that the value of $1\ R_{\rm ISCO}=6\ R_{g}$.

\end{table*}

Spectral modeling is performed with {\sc xspec} \citep{arnaud96} v12.14.1 using C-stat with errors reported at the 90\% confidence level throughout. The spectra are considered in the 2.4--17.4 keV passband similarly to \cite{NGC4151} and optimally binned \citep{kb16} with no minimum counts per bin specified. Absorption along the line of sight is accounted for with {\sc tbabs} with column density fixed to $N_{H}=4.4\times10^{21}\ \rm cm^{-2}$ \citep{HI4PI}. The $t_{avg}$, LB, and UB spectra are modeled concurrently with parameters allowed to vary between them. 

We initially model the continuum using two thermal components to describe emission from the accretion disk ({\sc diskbb}) and near the NS surface ({\sc bbody}). This provided an adequate description of the data below 10 keV, but was not able to account properly for the harder X-ray emission above 12 keV. The total fit statistic was C-stat = 25990.38 for 13580 degrees of freedom (dof) and 12 free parameters. Individually, there are 4 free parameter in each spectrum. The fit statistic per spectrum was C-stat = 12908.01 ($t_{avg}$), 6500.14 (LB), and 6582.24 (UB), with 4779 bins, 4377 bins, and 4436 bins, respectively. We calculate the corresponding Bayes information criterion (BIC = C-stat$\rm +k \log(n)$, where k is the number of free parameters and n is the number of bins: \citealt{bic}) for a total BIC = 26104.58 for this model or individual BIC of 12941.90 ($t_{avg}$), 6533.68 (LB), and 6615.83 (UB). We add a power-law component (typically used in the ``hybrid model" \citep{Lin07} for NS LMXBs) to account for the weak hard X-ray tail. 
This provides a significant improvement in the overall fit of $\Delta$C-stat = 8265.02 for 6 additional degrees of freedom;  corresponding to a BIC of 17896.67 or $\rm \Delta BIC=8207.91$ in comparison to the fit without the power-law component (or $\rm \Delta BIC > 1000$ for each spectrum). See Table~\ref{tab:spec} for parameter values. 
We alternatively tried a physically motivated description of Comptonization using the convolution kernal {\sc thcomp} \citep{thcomp} assuming that the seed photons originate from close to the NS (i.e., the {\sc bbody} component). We extend the energy range for model computation to 0.1 keV to 1000 keV using 1000 logarithmic bin, however, the use of {\sc thcomp} 
leads to unphysical values of the photon index that are extremely hard ($\Gamma\leq1.46$) with a very low electron temperature ($3.3\leq kT_{e}\ (\rm keV)\leq4.3$) and covering fraction that tends to small values ($<0.1$), which essentially indicates the the model tends to the original seed photons of the blackbody.  
The overall fit worsens by $\Delta$C-stat = 146.31 for 3 additional degrees of freedom; corresponding to a BIC of 18071.53. In comparison to the simpler continuum description using a power-law component, the use of {\sc thcomp} is strongly statistically disfavored in this case. We defer further testing of additional Comptonization models to a subsequent analysis that includes additional data from \nustar, which are not considered here as the cross-calibration of \xrism with other missions is still under investigation by the \xrism calibration team.

\begin{figure*}[t]
 \begin{center}
  \includegraphics[width=0.98\textwidth]{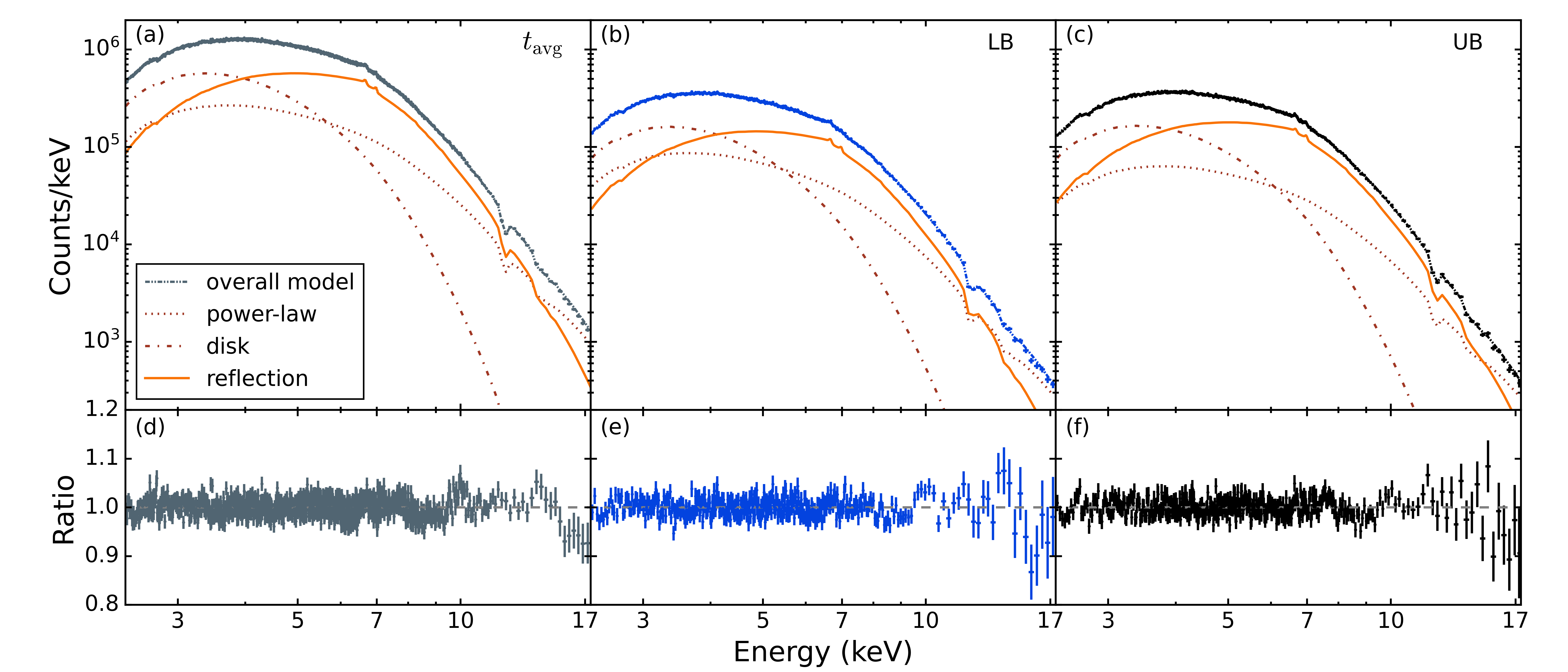} 
  \includegraphics[width=0.98\textwidth]{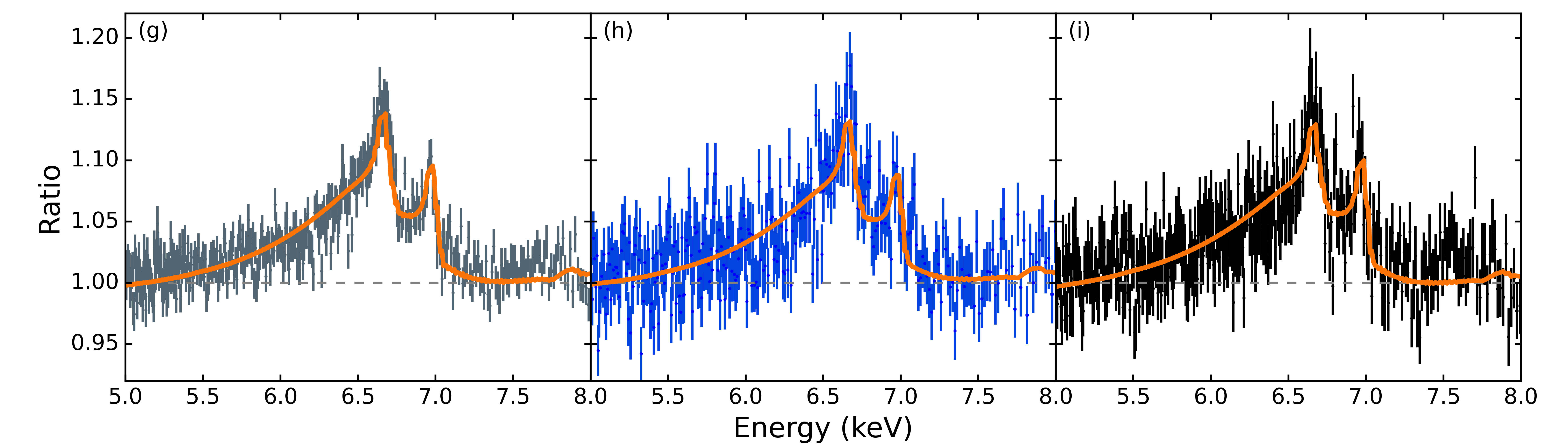} 
 \end{center}
\caption{The \xrism/Resolve counts spectrum and model components for (a) the time-averaged ($t_{\rm avg}$), (b) lower banana (LB), and (c) upper banana (UB) states, respectively. The overall model is indicated by the dot-dot-dot-dashed line, the power-law by the dotted line, the multi-temperature disk blackbody by the dot-dashed line, and the reflection model by the solid line. The panels (d)--(e) directly below the counts spectra show the ratio of the data to the overall model for each spectrum. Panels (g)--(i) show the Fe line profile from Figure~\ref{fig:fe} with the reflection model line predicted profile overlaid based upon fitting the counts spectra. 
Data were rebinned for clarity. 
}\label{fig:spec}
\end{figure*}

With a better continuum characterization, emission structure is revealed in the Fe~K band that is similar to the relativistic reflection inferred at low resolution (Figure~\ref{fig:fe}).  We therefore utilized \relxillns \citep{Garcia22}, which self-consistently calculates the reflection spectrum for irradiation of the accretion disk by a blackbody component.  We allow the reflection fraction ($f_{\rm refl}$) to take on positive values to maintain self-consistency between the illumination and reprocessed emission, thus the illuminating thermal component is included within the model and {\sc bbody} model component is removed from the overall fit. The dimensionless spin is fixed at $a=0$ and the outer disk radius is set at 1000 \rg. The inclination, $i$, iron abundance, $A_{\rm Fe}$, and disk density, $\log(n_{e}/\rm cm^{-3})$, are tied between the spectra since we do not expect these parameters to change as the source varies in intensity. The outer and inner emissivity index are tied together to create a single illumination profile. The emissivity parameter, $q$, inner disk radius, \rin, ionization state of the material, $\log(\xi)$, blackbody temperature, $kT_{bb}$, reflection fraction, $f_{\rm refl}$, and model normalization are free to vary between spectra. The addition of the reflection model significantly improves the overall fit ($\Delta$ C-stat = 3606.8 for an additional 15 dof, BIC = 14432.63, $\rm \Delta BIC=3464.04$ in comparison to the continuum model or $\rm \Delta BIC > 500$ for each spectrum) and provides an adequate description of the data, particularly in the Fe line region (Figure~\ref{fig:spec}). The results can be seen in Table~\ref{tab:spec}. 

The best fit parameter values do not differ significantly between the LB and UB. The spectra and model components are shown in Figure~\ref{fig:spec} as well as the ratio of the model fit to the data.  The disk is moderately ionized and viewed at a low inclination of $i=5^{\circ}\pm1^{\circ}$ with the inner radius remaining close to the innermost stable circular orbit in each case. 
We note that the iron abundance is $\sim3.3\times$ solar. Supersolar Fe abundances have been reported in the literature previously for \source \citep{miller13, ludlam18, mondal20, hall25} and a number of other Galactic X-ray binaries (see \citealt{ludlam24} for discussion and references therein), which can be attributed to degeneracy with the disk density in the reflection model \citep{tomsick18, garcia18}. We check how this impacts the \rin and inclination by fixing $A_{\rm Fe}$ at unity. The overall fit significantly worsens (total C-stat = 14269.39: $\Delta$C-stat = 150.83 increase for 1 degree of freedom) as the reflection model produces a weaker \ion{Fe}{26} emission line, but the inclination remains at $i=5^{\circ}\pm1^{\circ}$ and \rin is consistent within the 90\% confidence level of Table~\ref{tab:spec}, albeit with larger uncertainty ($t_{\rm avg}$: \rin = $1.23_{-0.21}^{+0.19}$ \risco, LB: \rin = $1.13_{-0.10}^{+0.96}$ \risco, UB: \rin = $1.30_{-0.25}^{+0.77}$ \risco). This indicates that the inferred supersolar Fe abundance does not strongly bias the reflection modeling results. 

Lastly, we determine the emission radii from the disk and blackbody components and compare to the values inferred from the reflection model. Using the source distance of $7.7\pm0.9$ kpc, a standard color correction factor of 1.7 \citep{shimura95}, and inclination of $i=5^{\circ}\pm1^{\circ}$, we find an inner disk radius between 17.5--31.97 km  from the normalization of the {\sc diskbb} component when propagating errors for each value. Since the single-temperature blackbody is included in the reflection modeling, we need to decompose the model into the input spectrum and reprocessed emission. To do so, we freeze the reflection refraction at the negative value of the best fit reported value in Table~\ref{tab:spec} and add a {\sc bbody} spectral component to the overall model with temperature fixed to the value of the illuminating component in the reflection model. We freeze all other parameter values in the model and fit the spectrum to only obtain the normalization of the {\sc bbody} component and find values of norm$_{\rm bb}$ ($10^{-2}$) = $1.784\pm 0.004$ ($t_{\rm avg}$), norm$_{\rm bb}$ ($10^{-2}$) = $1.45\pm0.01$ (LB), and norm$_{\rm bb}$ ($10^{-2}$) = $2.22\pm0.01$ (UB). Using these normalizations, the blackbody temperature of the reflection model in Table~\ref{tab:spec}, as well as aforementioned distance and color correction factor, the spherical emission radius ranges between 6.3--9.2~km. This is smaller than the anticipated radius of a NS. If the emission instead arises from a narrow banded region with a height that is 10\% the radial extent as expected for boundary layer emission \citep{popham2001}, then then emission radius of the blackbody component increases to 19.9--28.9~km, which is consistent with the inference of the inner disk from the {\sc diskbb} component. Assuming a canonical NS mass of 1.4 \ms, the inner disk radius inferred from the reflection model ranges from 12.5--19.3~km. While broadly consistent with the inferred radii of the disk and blackbody, the exact values depend on the mass of the NS which is currently unknown. A larger NS mass increases this estimate from the reflection model, whereas varying the assumed color correction factor changes the inferred radii from the thermal components (e.g., see \citealt{kubota01}). However, the general agreement between the inferred radii from the thermal components and reflection model suggests an extended boundary layer region is illuminating the accretion disk. 

% Anderson-Darling
% double thermal: log(Anderson-Darling)        -10.94     using 13592 bins.
% double thermal+pow: log(Anderson-Darling)        -12.94     using 13592 bins.
% reflection: log(Anderson-Darling)        -16.13     using 13592 bins.

% BIC = CSTAT + k*ln(n): k= # of free parameters, n = number of bins
% double thermal (xr_diskbbody_3spec_04032025.xcm): C-stat= 25990.38, bins=13592, k= 12: BIC = 26104.58
% double thermal+pow (xr_diskbbodypl_3spec_04032025.xcm): C-stat= 17725.36, bins=13592, k= 18: BIC = 17896.67
% Reflection (xr_diskplrelxillns_3spec_04032025.xcm): C-stat= 14118.56, bins=13592, k= 33: BIC = 14432.63
% Reflection+distant reflector (xr_diskbbodyplrelxillns_xillverns_3spec_07242025.xcm) = 14041.49, bins=13592, k=39, BIC = 14412.66 
% Reflection + gabs (gabsrelxillnstestv2_3spec_07312025.xcm):  C-stat= 14352.75, k = 36, BIC = 14695.37
% Reflection + gabs + 2 gaussian absorbers (gabs2gaussrelxillnstest_3spec_08052025.xcm): C-stat = 14290.02, k = 39, BIC = 14661.19
% reflection+warmabs (warmabs_test_07072025.xcm)= c-stat = 14110.89, bins=13592, k=36: BIC= 14453.51
% reflection + apec (apecrelxillns_08082025.xcm) : C-stat: 14087.85, k=39, BIC = 14459.02
% reflection + bapec (bapecrelxillns_08082025.xcm): C-stat = 13914.66, k=40, BIC = 14295.35

\subsection{Exploring Alternatives for the Observed Fe Line Profile }

The shape of the Fe line in \source resembles the recently reported emission line in GX 340+0 observed with \xrism/Resolve \citep{ludlam25, chakraborty25}. However, in the case of GX~340+0, the structure in the Fe~K band could not be explained with reflection alone. Alternative scenarios for the observed Fe line profile include reduced Fe abundance in the interstellar medium (ISM) along the line of sight \citep{dai09}, absorption from an outflowing wind \citep{dai09, cackett10, miller16, chakraborty25}, or additional emission from an ionized emitter farther out in the system \citep{ludlam25, chakraborty25}. We explored these alternative scenarios for the shape of the Fe line observed in \source. Given that the spectral parameters between the LB and UB do not change significantly, we focused the tests on the time-averaged spectrum.

We started by exploring the abundance of Fe in the ISM. We replaced {\sc tbabs} in the continuum model with {\sc tbfeo}, which allows for variable abundance of Fe and O.  We set the abundance of Fe to the minimum value of 0.5, while leaving the abundance of O fixed at unity (since we are insensitive to absorption by O which occurs below the \xrism energy band while the gate valve is closed).  The resulting fit is worse by $\Delta$C-stat = 45.6 for the same number of degrees of freedom ($\Delta \rm BIC=-45.6$; indicating no evidence for reduced Fe abundance).  This indicates that the dual-peaked profile is not an artifact of modeling the Fe K edge poorly.  This is more readily discerned in \source because its ISM column is at least an order of magnitude lower than the column along the line of sight to GX~340+0 (which is $\mathcal{O}(10^{22-23}\ \rm cm^{-2})$: \citealt{cackett10, zeegers19, corrales25}).

\begin{figure*}[t]
 \begin{center}
  \includegraphics[width=0.98\textwidth, trim= 20 20 0 0, clip]{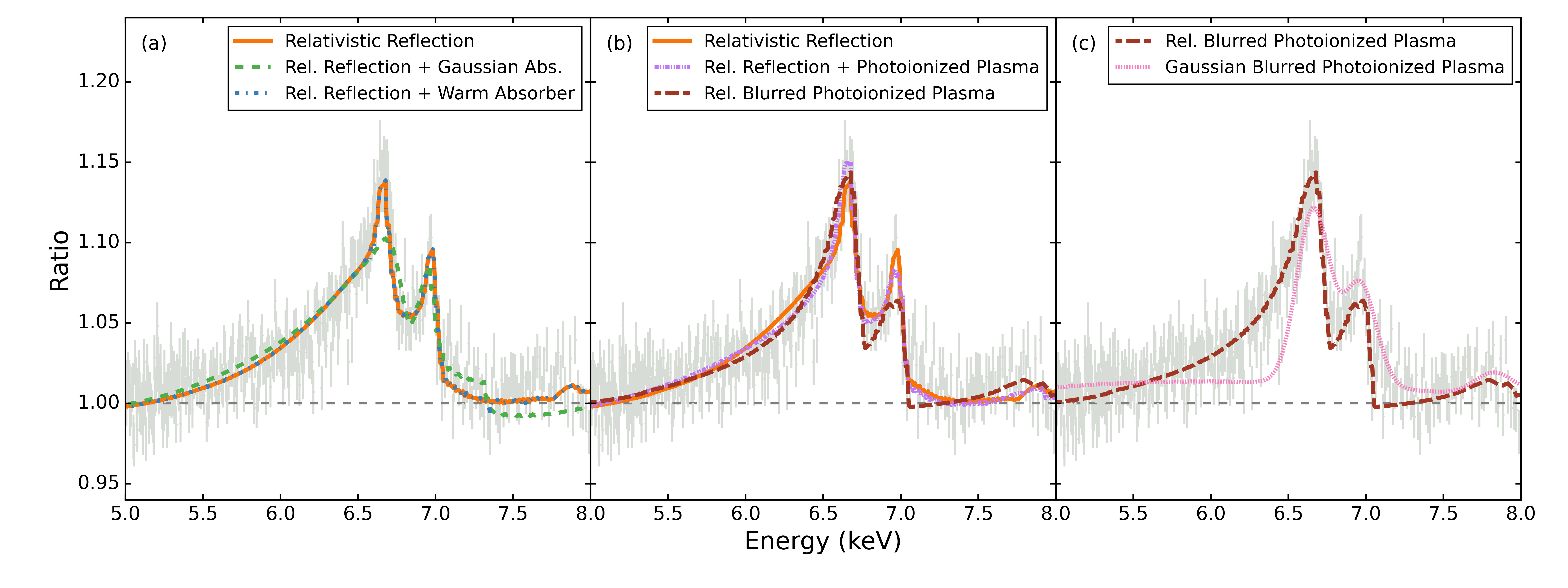} 
 \end{center}
\caption{Predicted Fe line profile from the use of (a) absorption, (b) photoionized emission models, and (c) different blurring kernals. Time averaged data are shown in light grey and were rebinned for plotting purposes. The reflection (\relxillns) model predicted line profile is shown as a solid orange line in panels (a) and (b). In panel (a) the green dashed line indicates the model predicted line profile when a Gaussian absorber is required, whereas the blue dot-dashed line shows the line profile when {\sc warmabs} is applied (the profile of which overlaps with the reflection only model prediction). In panel (b) the purple dot-dot-dot-dashed line shows the line profile with the addition of a photoionized emitting plasma in the system. The \ion{Fe}{25} line is stronger and broader in comparison to the reflection model predicted line profile, but the \ion{Fe}{26} line is less pronounced. The red long dashed line in panel (b) denotes the line predicted model profile after removing the reflection model and allowing relativistically blurred photoionized emission to fit the line complex. The model appears to under-predict the \ion{Fe}{26} emission. In panel (c) we show the photoionized emission component convolved with relativistic effect (red dashed line) versus a Gaussian blurring kernal (pink dotted line); the latter of which is unable to adequately describe the red-wing of the line profile. 
}\label{fig:mocomp}
\end{figure*}

Next we tested for the presence of an absorber superimposed upon the emission line. We applied a simple Gaussian absorption line ({\sc gabs}) between 6.7--6.97 keV to the overall fit with relxillNS. A line centroid of $6.83\pm0.01$ keV with a width of $\sigma = 57.9_{-6}^{+4}$ eV and depth of $\tau = 0.0013\pm0.0002$ indicative of an optically thin region, could be fit to the dip between the \ion{Fe}{25} and \ion{Fe}{26} emission lines. As pointed out in \cite{miller16} for a similar line in GX~340+0, this line energy would either correspond to a strong inflow of \ion{Fe}{26} or a strong outflow of \ion{Fe}{25} with a velocity of $v\simeq0.02c$ in either case.  However, requiring the superposition of absorption on the reflection model resulted in a worse overall fit (C-stat/dof = 5352.09/4763 or BIC = 5487.64) in comparison to the fit using the reflection model alone (see Table~\ref{tab:spec}). 

Moreover, if an absorber were present to cause the strong dual-peaked line profile, there should be multiple absorption features apparent within the spectra. We searched for additional absorption features by applying two Gaussian lines with a fixed width of 50 eV \citep{cackett10}, separated by an energy of 0.3 keV in the higher energy band of 7.8 -- 9.3 keV where the \ion{Fe}{25} and \ion{Fe}{26} beta lines are expected. The addition of the two absorption lines improved the fit in comparison to applying {\sc gabs} alone (BIC = 5477.07 or $\Delta \rm BIC = 10.6$) for lines at $8.55_{-0.08}^{+0.03}$~keV and $8.85_{-0.08}^{+0.03}$~keV with an equivalent width of $\sim5$~eV and $\sim4$~eV, respectively. These line energies imply an even stronger outflow than the {\sc gabs} model with $v>0.06 c$. The inner disk radius becomes larger (e.g., for the time averaged spectrum \rin $=1.41_{-0.04}^{+0.06}$~\risco) and the inclination of the system is higher ($\sim30^{\circ}$), however, the overall fit is still not statistically preferred over the reflection model alone. We show the overall model predicted Fe K emission line profile in Figure~\ref{fig:mocomp}(a) in comparison to \relxillns. 

Given that there should be an array of transitions from atomic elements with various expected line intensities all shifted by the same velocity if an outflow were present, we removed the simplistic Gaussian absorption lines and utilized {\sc warmabs}\footnote{\href{https://heasarc.gsfc.nasa.gov/xstar/docs/sphinx/xstardoc/docs/build/html/index.html}{See {\sc warmabs} documentation here.}} (v2.49d) based off of {\sc xstar} \citep{kallman01}. The model treats the irradiation of an optically thin medium by a power-law continuum with $\Gamma=2$, which is approximately the shape of the emission within the region of concern for \source and sufficient for testing the proposed scenario. We assumed a turbulent velocity of $\sigma = 300~{\rm km}~{\rm s}^{-1}$ and solar abundances. We allowed the column density, ionization parameter, and redshift to be free parameters. The column density of $(5.29\pm0.01) \times10^{20}$ cm$^{-2}$ is an order of magnitude lower than the absorption from the ISM, causing the fit to be insensitive to the ionization state of the absorber. The absorber is blue-shifted at a velocity of $v=0.02c$. The inferred inner disk radius and inclination are consistent with 90\% confidence level as reported in  Table~\ref{tab:spec} indicating no impact on key parameters with the addition of {\sc warmabs}. The overall fit is improved (C-stat/dof = 5200.76/4763 or BIC = 5336.31) in comparison to using individual absorption lines, however, when compared to the reflection only fit of Table~\ref{tab:spec} the addition of {\sc warmabs} is statistically disfavored and has no discernible impact on the observed Fe emission line profile (see Figure~\ref{fig:mocomp}(a)). Consequently, there is no strong evidence for the presence of an absorber in \source that would be responsible for the observed dual-peaked line profile.

\begin{table*}[!t]

\caption{Comparison of spectral models and change in key parameters for the time-averaged spectrum}
\label{tab:comp} 

\begin{center}

\begin{tabular}{lccccccc}
\hline
Model & \rin (\risco) & $i$ ($^{\circ}$) & C-stat & dof & C-stat/dof &  BIC & $\Delta$BIC \\
\hline
\relxillns & $1.02_{-0.01}^{+0.21}$ & $5\pm1$ & 5205.64 & 4766 & 1.09 & 5315.78 & --\\
+ 3 Gaussian Absorbers & $1.41_{-0.04}^{+0.06}$ & $30\pm1$  & 5316.10 & 4760 & 1.12 & 5477.07 & $-161.29$\\
+ {\sc warmabs} & $1.05\pm0.01$ & $5\pm1$ & 5200.76 & 4763 & 1.09 & 5336.31 & $-20.53$ \\
+ {\sc apec} & $1.06_{-0.06}^{+0.14}$ & $4\pm1$& 5188.68 & 4764 & 1.09 & 5315.76 &$+0.02$\\
+ {\sc bapec} & $1.13_{-0.12}^{+0.09}$ & $4_{-1}^{+2}$ & 5083.74 & 4763 & 1.07 & 5219.29 & $+96.49$\\
+ {\sc photemis} & $1.24\pm0.01$ & $8\pm1$ & 5134.42  & 4762 & 1.08 & 5278.44 & $+37.34$ \\
Rel.\ blurred {\sc photemis} & $1.09_{-0.02}^{+0.03}$ & $11\pm1$ & 5433.88 & 4768 & 1.14 & 5527.07 & $-211.294$ \\

\hline

\end{tabular}
\end{center}

\medskip
Note.---  Errors are reported at the 90\% confidence level. \rin is quoted for the time-averaged spectra. $\Delta$BIC is in comparison to the reflection model fit statistic in Table~\ref{tab:spec} for the time-averaged spectrum only, where a negative value indicates a worse fit and no evidence in support of the additional component, whereas a positive value indicates evidence in favor. For reference, there are 4779 bins present in the time-averaged spectrum.

\end{table*}

We next considered whether a collisionally ionized plasma ({\sc apec}: \citealt{Smith01}) may contribute to the spectrum or even dominate over the reflection component. We allowed the temperature and normalization to vary, but fixed the elemental abundance to solar values.  The temperature of the plasma exceeds that of the disk and boundary layer, indicating an energetic distribution of electrons; however, the fit is statistically worse and not preferred over the reflection model alone (C-stat/dof~=~5188.68/4764 or BIC~=~5315.76) though the inner disk radius and inclination remain consistent within the 90\% confidence level with Table~\ref{tab:spec}. Allowing the plasma to be velocity broadened (i.e., {\sc bapec}) provided an improvement in the overall fit (C-stat/dof = 5083.74/4763 or BIC = 5219.29) with no notable change in key parameters, but requires a large velocity value of $v=2388_{-321}^{+493}$ km/s. Since this is a projected velocity, the deprojection at the low inferred inclination of $5^{\circ}$ would imply that the plasma is located $\sim120$~\rg assuming Keplerian motion. If this were to arise from the corona of the accretion disk, we estimate the density of the medium from the normalization of the model ($K=0.12\pm0.03$ for the time-averaged spectrum) assuming an extended thin disk atmosphere (with $H = 0.01 R$) of uniform density gives $n_{e}\simeq1.3\times10^{19}\ \rm cm^{-3}$, which is denser than the disk inferred from reflection but within expectations for the density of the inner disk around NSs \citep{Frank02}. If the geometry of the medium were spherical, this reduces the density by an order of magnitude to $n_{e}\simeq1.3\times10^{18}\ \rm cm^{-3}$, but would obscure the line of sight to the central source \citep{Jimenez02, schulz09}. Thus, while the model provides a statistical improvement, physically the parameter values are not viable even though it does hint at evidence of additional emission present in the system.  

Finally, we examined the potential influence of a photoionized emitter ({\sc photemis} v2.49d). Again, we assumed a turbulent velocity of $\sigma = 300~{\rm km}~{\rm s}^{-1}$ and solar abundances, and allowed the ionization and normalization to vary. Given that the collisionally ionized plasma properties indicated emission within 1000 \rg of the NS and thus co-located with the reflected emission, we fix the redshift parameter to 0 and apply the blurring kernal {\sc relconv} \citep{Dauser10} to account for relativistic effects imparted to the photoionized emission. The addition of the photoionized plasma to the reflection model provided an improvement to the overall fit (C-stat = 5134.42 or BIC = 5278.44). Again, the key systemic parameters did not change significantly, but the inclination of the system increased slightly to $8^{\circ}\pm1^{\circ}$ and \rin increased to the upper end of the 90\% confidence level from the reflection-only fits. The inferred position of the photoionzed plasma is closer to the source ($\sim33$ \rg) than in the case of the collisionally ionized emission. 

The model-predicted line profile with the addition of {\sc photemis} to the reflection model is shown in Figure~\ref{fig:mocomp}(b) with the reflection predicted line profile provided for comparison. The overall model with photoionized emission predicts a stronger \ion{Fe}{25} emission line but weakens the contribution from \ion{Fe}{26}.
Given the co-location of the emission from the photoionized plasma and reflection, we test the most extreme case where the observed line profile could be caused by a run of transition lines from an ionized plasma and remove the reflection from the overall spectral model. In this case, the fit statistic significantly increases (C-stat = 5433.88 and BIC = 5527.07) indicating that this scenario is not statistically preferred over reflection. The location of the photoionized plasma must be close to the NS (\rin $=1.09_{-0.02}^{+0.03}$) in order to capture the shape of the red-wing in the Fe line profile (see Figure~\ref{fig:mocomp}(b)) at a low inclination, but the model is unable to capture the relative strength of the \ion{Fe}{26} emission observed in the data. We summarize the fit statistic and impact on key parameters like the inner disk radius and inclination in Table~\ref{tab:comp}.

Based on Figure~\ref{fig:mocomp}(b) and the marginal change in key parameters in Table ~\ref{tab:comp} inferred from the various tests (including relativistic effects convolved with different emission models), it would appear that the atomic physics of the  models themselves may not be as important as the relativistic effects as long as they can relatively predict the strongest lines observed within the spectrum. Similar results are found in the literature when using different disk reflection models blurred with relativistic convolutions (e.g., \citealt{Garcia22}). We therefore check if the Fe line structure can be adequately described without requiring relativistic effects at all.
To test the necessity of the relativistic blurring, 
we remove {\sc relconv} and apply a Gaussian smoothing kernal ({\sc gsmooth}) to the photoionized emission component instead. Linear energy bins of 5~eV are set via the `energies' command in {\sc xspec} as necessitated by the model. A uniform Gaussian width of $\Sigma=0.10\pm0.01$ keV is necessary to broaden the Fe emission lines in {\sc photemis} to approximately the shape observed in the data. However, this convolution is unable to adequately describe the asymmetry (i.e., the red-wing) observed within the structure the Fe line complex (see Figure~\ref{fig:mocomp}(c)). The overall fit is significantly worse (C-stat = 6210.5 or BIC = 6278.3) in comparison to when relativistic blurring was applied to the photoionized emission. The same test can be applied with the reflection only component ({\sc xillverNS}) of the relativistic reflection model. The disk inclination, iron abundance, ionization, and disk density is fixed to the values inferred from the relativistic reflection model. In this case, a uniform Gaussian width of $\Sigma=0.07\pm0.01$ keV is needed to broaden the rest-frame reflected emission, but the fit is again statistically worse (C-stat = 5548.5 and BIC = 5616.26). In both cases, the difference in the BIC between Gaussian blurred emission and relativistically blurred emission exceeds 300 (i.e., $\rm BIC_{gsmooth} - BIC_{relconv} > 300$). This strongly disfavors Gaussian blurring over relativistic blurring. Ultimately, given the different scenarios explored, 
relativistic reflection remains the simplest model to describe the shape of the Fe line structure in \source though additional emission from ionized plasma cannot be ruled out. 

\section{Discussion} \label{sec:discussion}

We present the first \xrism/Resolve observation of the NS LMXB \source. During the 171 ks exposure, the source varied between the lower and upper banana branch, thus remaining in a soft spectral state throughout. Inspection of the time averaged and flux-selected data revealed a broad asymmetric Fe~K emission line profile that contained structure indicative of H-like and He-like Fe K-alpha transitions. Spectral modeling indicates that source parameters do not vary significantly between the upper and lower banana branch. Despite the conflicting reports in the literature, the observed spectrum is most successfully described in terms of relativistic reflection arising from the inner accretion disk.  The best-fit model for the time-averaged spectrum gives an inner radius of \rin~$=1.02_{-0.01}^{+0.21}$~\risco (assuming $a=0$), and an inclination of $5^{\circ}\pm1^{\circ}$.  Potential alternative explanations of the structure in the Fe~K band, including anomalous abundances along the line of sight and intrinsically narrow (or, highly broadened) collisional or photoionized emission fail to match the data or modestly change the best-fit reflection parameters.  Furthermore, relativistic effects are necessary to properly account for the asymmetric line profile observed. 

The XRISM/Resolve spectrum of Cygnus X-1 also reveals relativistic reflection from the inner accretion disk, and measures a spin parameter consistent with values reported based on NuSTAR data (Draghis et al.\ 2025, $submitted$).  At the opposite end of the mass scale, the XRISM/Resolve spectra of NGC~4151 and NGC~3783 appear to prefer very broad components at the base of their Fe~K emission line complexes (\citealt{xrism2024}, \citealt{mehdipour2025}).  In every class of objects wherein charge-coupled devices (and other solid-state imaging spectrometers) have detected relativistic emission lines,  Resolve appears to confirm these findings with broadly consistent results.  The special power of Resolve is that it is able to detect or rule-out viable alternatives that rely on complex absorption and/or collections of narrow emission lines.  

It is notable that high-resolution X-ray spectroscopy has confirmed that the inner disk in \source is likely viewed at a very low inclination, $5^{\circ}\pm 1^{\circ}$ (see Table~\ref{tab:spec}).  Given the fact that optical studies measure a similarly low inclination, the entire accretion flow in \source may be aligned.  This stands in contrast to some stellar-mass black holes in wide binary orbits with massive stars, such as Cygnus X-1 and V4641~Sgr.  The shorter orbital period of $P\sim 2$~hours in \source \citep{Cornelisse13} suggests a far more evolved binary system that has had time to come into alignment.  Indeed, the orbital period of \source is so short that the companion may be degenerate; the mass ratio of the components may also play a role in whether or not a binary system reaches alignment.  We note that the magnetic axis of the NS does not have to be aligned with its rotational axis, nor the angular momentum of the system.

Radio emission appears to be genuinely quenched in stellar-mass black holes in soft, disk--dominated states \citep{russell2011}.  In contrast, \source is sometimes detected in radio bands, despite the fact that it is persistently soft \citep{pattie2024}.  It is possible that this is also due to a low inclination: any radio emission from a jet would have a greater chance of being beamed into our line of sight, amplifying what might be an intrinsically weaker flux.   

The low inclination of \source may also be a key factor in revealing the relativistic content of the Fe~K line in this system.  This viewing angle limits the role of Doppler shifts, keeping the line flux relatively compact, and making it more distinct relative to the continuum.  Similarly, although disk winds may be launched vertically, most of the column density in disk winds appears to be concentrated equatorially, along the face of the disk (e.g., \citealt{miller2008}).  At a higher inclinations, even a strong relativistic line may only be seen after it is distorted by a strong disk wind.  Recent fits to the XRISM spectrum of GX~340$+$0, for instance, measure an inclination of $i = 39^{\circ}\pm 1^{\circ}$, and the relativistic line shape is partly distorted by plasma lines \citep{ludlam25}.   Future XRISM observations can test the role of disk winds, scattering, and other key effects by targeting NS LMXBs with that span a range of established inclination values.

This work has tested the relativistic reflection paradigm in NS LMXBs using high-resolution spectroscopy, but it has not harnessed the full power of the data.  In future work, we will explore the ability of these data to constrain the spin parameter of the NS, potentially enabling stronger constraints on the NS equation of state.  We will also examine the burst spectra in detail, and report those results in a third paper.
\\

{\it Acknowledgments:} 
This work was supported by NASA under grant No.\ 80NSSC25K7852.

\bibliography{references}

\end{document}